\documentstyle[aps,twocolumn,prl,epsf]{revtex}

\begin{document}

\newcommand{\be}{\begin{equation}}
\newcommand{\ee}{\end{equation}}
\newcommand{\bn}{\begin{eqnarray}}
\newcommand{\en}{\end{eqnarray}}
\newcommand{\bq}{\begin{eqnarray}}
\newcommand{\eq}{\end{eqnarray}}

\draft

\twocolumn[\hsize\textwidth\columnwidth\hsize\csname @twocolumnfalse\endcsname

\draft

\title{Towards a Fully Ab-Initio Description of the Diluted Magnetic 
Semiconductor $Ga_{1-x}Mn_{x}As$.  Ferromagnetism, Electronic Structure, 
and Optical Response.}

\author{L. Craco, M. S. Laad, and E. M\"uller-Hartmann }

\address{${}^1$Institut f\"ur Theoretische Physik, Universit\"at zu K\"oln, 
Z\"ulpicher Strasse, 50937 K\"oln, Germany}

\date{\today}

\maketitle

\widetext

\begin{abstract}
There are two competing views of itinerant ferromagnetism, the first 
viewing ferromagnetism as resulting from the indirect coupling between local
moments via the itinerant carrier dynamics, the so-called RKKY mechanism,
while in the alternative picture, ferromagnetism results from the spin
polarization of itinerant carriers by the strong atomic Hund interaction - 
the so-called double exchange (DE) scenario. Which view describes the 
ferromagnetism in diluted magnetic semiconductors, materials with promise 
for spintronic applications, is still unclear. Here, we describe the detailed 
physical response of the prototype material $Ga_{1-x}Mn_{x}As$ using a 
combination of first-principles bandstructure with methods based on dynamical 
mean field theory to incorporate strong, dynamical correlations {\it and} 
intrinsic as well as extrinsic disorder in one single theoretical picture.  
We show how ferromagnetism is driven by DE, in agreement with very recent 
observations, along with a good quantitative description of the details 
of the electronic structure, as probed by scanning tunnelling microscopy 
(STM) and optical conductivity.  Our results show how ferromagnetism can 
be driven by DE even in diluted magnetic semiconductors with small carrier 
concentration. 
\end{abstract}


]     

\narrowtext

Spintronics is a rapidly emerging technology where it is not the electron 
charge but the electron spin that carries information.  It offers promising
opportunities for developing a new generation of devices based on a 
combination of standard microelectronics with spin-dependent effects arising 
from the interaction of the carrier spin with the magnetism of the 
material~\cite{[1]}.  Another promising area envisions its marriage with 
optical photons, with the possibility of designing new spin based devices 
such as spin-field effect transistor (FET), spin-light emitting diodes (LED), 
optical switches operating in the THz range, modulators, encoders and 
decoders, and bits for quantum computation~\cite{[2]}.  
Discovery of optically induced ferromagnetism in some $Mn$-doped III-V
semiconductors opens up the possibility of applications to photonic storage 
devices and photonically-driven micromechanical elements~\cite{[1]}.
Success of this vision demands a deeper understanding of fundamental
spin interactions in the solid state, along with the realistic bandstructures,
and roles of dimensionality as well as intrinsic (extrinsic) defects.

The discovery of dilute magnetic semiconductors (DMS), which are III-V 
semiconductors randomly doped with small amounts of magnetic atoms, like 
$Mn^{2+}$, has provided us with an attractive example of prototype materials 
of great interest in this context~\cite{[3]}. These have ferromagnetic 
transition temperatures much higher than those of earlier known 
$Eu$-chalcogenides~\cite{[3]}, of order $100~K$, and the magnetism can be 
controlled electronically. The optimum value of $x$ in $Ga_{1-x}Mn_{x}As$ 
corresponding to highest $T_{c}=100~K$ is $0.043-0.05$.  In spite of much 
activity, the nature of ferromagnetism, as well as the detailed electronic 
structure, is not properly understood. Ultimately, the mechanism of FM 
should go hand-in-hand with the details of spin interactions in a system 
of dilute $Mn^{2+}$ ions doped randomly in the $GaAs$ host, and to the 
details of modification of the electronic structure due to $Mn$ doping.

Application of the Ruderman-Kittel-Kasuya-Yosida (RKKY) mechanism  
provides some understanding of magnetic and transport properties of 
DMS~\cite{[4]}, but is somewhat questionable in this case, since it is only 
valid when the exchange interaction, $J<<E_{F}$, the Fermi energy of the 
carriers.  This is obviously not the case in DMS, where the carrier 
concentration is low~\cite{[5]}.  An alternative double-exchange (DE) model 
based approach~\cite{[5]} has recently been invoked within a model 
bandstructure to study the evolution of $T_{c}$ with $x$.  However, the 
specific conditions under which DE ideas can be applied to DMS remain to be
elucidated. 

Here, we discuss an implementation of DMFT which permits the incorporation of 
the actual electronic structure of real materials into the DMFT equations.
Using this method, we show how a detailed understanding of half-metallic
ferromagnetism, along with a quantitative description of the correlated 
electronic structure of $Mn$-doped $GaAs$ is achieved.

$GaAs$ is a well known band semiconductor.  Its electronic structure, however,
is sensitive to $As$ antisites forming in the bulk during growth. The 
importance of these defects in the $GaAs$ host, constituting a source of 
{\it intrinsic} disorder, will become clear below. Each $Mn$ ion in $GaAs$ 
serves a dual purpose, acting as an acceptor as well as a magnetic impurity.  
In reality, the situation is a bit more complex when the realistic local 
characteristics of the $Mn$ ion, such as the charge transfer energy 
($E_{Mn}-E_{As}$), and the $p-d$ hybridization ($As-Mn$), and Coulomb 
 interactions, are taken into account.  The importance of including 
such quantum chemical aspects is shown very clearly in spectroscopic 
measurements~\cite{[6]}; these yield the on-site Hubbard $U=3.5~eV$, the 
charge transfer energy $\Delta=-1.5~eV$, and the hybridization 
$t_{pd}=1.1~eV$. Additionally, the Hund's rule coupling $J_{H}\simeq 0.55~eV$,
acting like a magnetic impurity potential~\cite{[5]}.  Finally, the 
random distribution of the $Mn$ ions gives rise to an additional non-magnetic 
disorder component.  All previous theoretical works done for $Ga_{1-x}Mn_{x}As$
have focused either on pure bandstructure aspects~\cite{[10]}, or on model
approaches~\cite{[5]}, and we are not aware of calculations which incorporate
the strong correlation at $Mn$ sites, along with local quantum chemical
information ($t_{pd},\Delta$) and {\it intrinsic} $As$ antisite disorder as 
well as $Mn$-doping related magnetic and non-magnetic disorder into the 
complex bandstructure of this material in a consistent way. Very recent 
theoretical advances~\cite{[11]} have opened an attractive possibility to 
study the correlated bandstructure of materials involving transition metal 
oxides and rare-earth based compounds via LDA+DMFT. 

In what follows, we show how a consistent theoretical (via LDA+DMFT) 
implementation including real bandstructure of $GaAs$, quantum chemical and
strong correlation 
aspects mentioned above, and $As$ antisite disorder is indeed necessary for 
a detailed quantitative understanding of the physics of $Ga_{1-x}Mn_{x}As$. 

A beautiful recent time-resolved magneto-optic measurement clearly shows 
the half-metallic character of the ferromagnetic metallic state in 
$Ga_{1-x}Mn_{x}As$~\cite{[7]}.  But this is not all, as the correlated 
nature of this state is further revealed by various responses.  First, 
photoemission measurements~\cite{[8]} reveal very small spectral weight near 
the Fermi surface, and a $T$-dependent build-up of spectral weight at higher 
energies. Further, the change of the chemical potential $E_{F}$ with $Mn$
doping is intriguing.  At small $x$, it appears to be almost pinned to its 
$x=0$ value, but starts moving towards the valence band (VB) around the $x$ 
value where the semiconductor-metal transition accompanied by FM occurs.  
At $x>0.05$, it moves up again, concomitant with occurence of insulating 
behavior with reduction of $T_{c}$, clearly showing the intimate connection 
between changes in electronic structure and variation in $T_{c}$ as a function 
of $x$. More anomalous features are observed in optical 
measurements~\cite{[9]}: there is no quasicoherent Drude response; instead, 
the spectrum exhibits a peak centered around $1200~cm^{-1}$, with a curious 
bump at lower energies.  From a ``modified Drude fit'', an effective mass 
$m^{*}=0.72m$ is deduced, in contrast to $m_{b}=0.24m$ from LDA calculations.  
The depression of the low-energy incoherent spectral weight with increasing 
$T$ (decreasing magnetization) is also very clearly seen; what is interesting 
is that this spectral weight transfer (SWT) takes place over a wide energy 
scale of about $4~eV$ (much larger than $T_{c}$). These observations are 
inexplicable within bandstructure ideas, and constitute  
direct evidence for the importance of dynamical electronic correlations in 
$Ga_{1-x}Mn_{x}As$.

\begin{figure}[htb]
\epsfxsize=3.5in
\epsffile{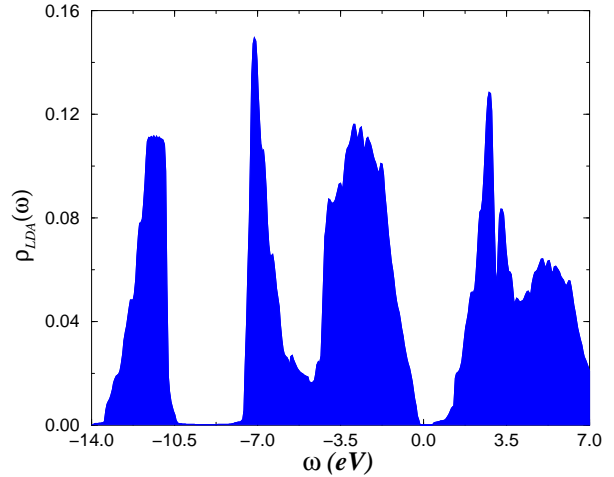}
\caption{Total density of states (DOS) of pure (without antisites) $GaAs$
computed within the Kohn-Sham local density approximation~[10]. 
The Fermi level is located at $\omega=0.0$.} 
\label{fig1}
\end{figure}

While it provides a good description of ground state properties, the local 
density approximation (LDA) cannot describe the excited states of correlated 
systems, since the dynamical correlations in a {\it quantum} many-body system 
are neglected in the LDA. Thus, given the above arguments, a realistic 
description of $Ga_{1-x}Mn_{x}As$ requires combining the LDA, which provides 
an excellent ab-initio one-electron bandstructure of weakly correlated metals 
and band insulators~\cite{[10],[11]}, with well controlled, state-of-the-art 
many body calculations capable of accessing the dynamical effects of strong 
electron correlations in a proper way.  Recently, the combination of LDA with 
dynamical mean field theory (DMFT) has been shown to provide good quantitative 
description for various correlated systems~\cite{[12]}.  Using LDA+DMFT, we 
will show how good quantitative agreement with DE ferromagnetism, STM data, 
and optical conductivity is obtained.   

Local atomic-like features of $Mn^{2+}$ in $GaAs$ are obtained from 
the set of many-body levels describing processes of electron addition and/or
removal in the atomic $d$-shell.  The constants $U, J_{H}$ can be computed from
first principles.  Itinerant aspects of the actual solid enter via a dynamical
bath function describing $p-d$ hybridization; it quantifies the
degree of itinerance of the $d$ electrons, and needs to be determined in a 
selfconsistent way from the DMFT equations.  With this, the problem is reduced
to solving an (asymmetric, in general) multi-orbital Anderson impurity 
problem selfconsistently embedded in a dynamical bath, giving us the impurity 
Green's function $G_{imp}(\omega)$ and the self-energy $\Sigma(\omega)$.

The bands of $GaAs$ are obtained from the eigenvalues of the matrix 
one-electron Hamiltonian $H(k)$, obtained from the LDA 
Kohn-Sham hamiltonian~\cite{[11]}. The actual LDA density of states (DOS) 
for $GaAs$ (without antisites) is shown in Fig.~\ref{fig1}.

To begin with, we model the effect of the random $As$ antisite potential in
pure $GaAs$ using the usual coherent potential approximation (CPA)~\cite{[13]}.
Next, given the small concentration of $Mn$ sites, we first solve the 
asymmetric Anderson impurity problem in the dynamical bath provided by the 
$GaAs$ bandstructure for a single $Mn$ impurity with the parameters given 
earlier, using the generalization of the iterated perturbation theory (IPT) 
for arbitrary filling ~\cite{[12]}. The exact low-frequency behavior is 
obtained from the Friedel-Luttinger sum rule~\cite{[12]}, while the correct 
high-frequency behavior is obtained from a selfconsistent computation of 
moments.  These equations have appeared previously 
in various contexts~\cite{[12]}, and we do not reproduce them here.
  For the one-band Hubbard model, very good agreement with quantum 
Monte Carlo data has been reported at high $T$.  For a concentration $x$ of 
$Mn$ impurities, one has additionally to perform a configurational average 
over random $Mn$ positions (non-magnetic) as well as an average over all 
spin configurations of the $Mn$ ions (magnetic). These effects of strong 
magnetic and positional disorder scattering are treated using the extended 
dynamical coherent potential approximation (CPA)~\cite{[5]}, which is 
combined with IPT for the $Mn$ impurities in a consistent way~\cite{[14]}.   

Marriage of LDA with DMFT gives us a quantitative description of {\it both}
ground- and excited state properties.  Finite temperature, and local moment 
effects above $T_{c}$ are readily handled using temperature Green's functions 
within DMFT.

\begin{figure}[htb]
\epsfxsize=3.5in
\epsffile{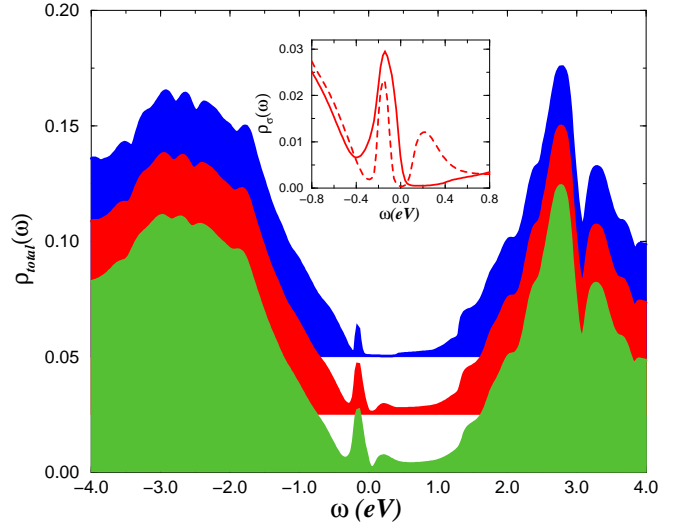}
\caption{Renormalized total DOS of $Ga_{1-x}Mn_{x}As$ as a function of $x$.
For $x=0$, the additional impurity feature at $\omega=-0.1$ is related to the
antisite ($n_{as}=0.015$) contribution (blue).  For $x=0.022$, a second 
$Mn$-doping related peak is resolved, but the system remains insulating 
(red).  With $x=0.043$ (green), half-metallic behavior is clearly seen in 
the inset (bold and dashed lines correspond to majority- and minority-spin 
DOS), corresponding to DE ferromagnetism. Notice the good agreement 
with STM data~[15].} 
\label{fig2}
\end{figure}

We now describe our results.  In Fig.~\ref{fig2}, we show the total density 
of states (DOS) for $Ga_{1-x}Mn_{x}As$ for $x=0, 0.022, 0.043$ for our chosen 
parameter set. Given~\cite{[15]} that each $Mn$ dopant creates 
an additional $As$ anti-site, the concentration of antisites, $n_{as}=0.015+x$.
For $x=0$, the effect of the random $As$ anti-site disorder is modeled 
using CPA, resulting in an impurity ($As$) band of anti-sites split off from
the VB in the semiconducting gap of $GaAs$.  We draw attention to the 
good agreement of the calculated ($x=0$) DOS with STM data~\cite{[15]}.  For 
small $x=0.022$, the total DOS still indicates semiconducting behavior, in 
coincedence with conclusions from 
Ref.~\cite{[16]}.  Notice, however, the appearance of a second, broader peak
in the gap.  Again, this is in accordance with results of STM 
measurements~\cite{[15]}.  Without random magnetic and non-magnetic impurity 
scattering, a quasicoherent peak is obtained (not shown).  Inclusion
of strong magnetic ($J_{H}$) and positional disorder scattering by CPA 
washes out this low-energy coherence, resulting in the red curve in 
Fig.~\ref{fig2}. For $x=0.043$, metallic behavior, together with 
ferromagnetism (see below), is clearly revealed in the 
results (Fig.~\ref{fig2}, green). From the real part of the one particle 
self-energy (not shown), we find 
$\partial\Sigma'(\omega)/\partial\omega|_{\omega=\mu}=-2.03$, leading to a 
mass enhancement of $m^{*}/m_{b}=3.03$ (here, $m_{b}$ is the carrier mass 
estimated from LDA), in excellent agreement with observations.  The 
half-metallic character of the system for $|\omega| \le 0.2$ is clearly seen 
from our computed results (inset to Fig.~\ref{fig2}), in full 
accord~\cite{[7]} with recent experimental work.

The ferromagnetic character of the metallic state is clearly shown by using
the DMFT propagators to compute the static part of the spin susceptibility,
$\chi_{s}({\bf q})=N^{-1}\sum_{\bf k}G_{\uparrow}({\bf k}+{\bf q})G_{\downarrow}({\bf k})$, and the effective intersite exchange, 
$J_{eff}({\bf q})=J^{2}\chi_{s}({\bf q})$.  Given the small concentration of 
the correlated $Mn$ ions, we expect that vertex corrections appearing 
normally in the computation of $\chi_{s}({\bf q})$ will not modify our 
estimate of $T_{c}$ much (these corrections scale with the $Mn$ concentration).
The transition temperature $T_{c}$ is estimated from the DMFT 
equations~\cite{[5]}.  We find, in full agreement with~\cite{[16]}, that 
$T_{c}=0$ for $x=0, 0.022$ and $T_{c}=120~K$ for $x=0.043$.

In light of our results, we see clearly that ferromagnetism in 
$Ga_{1-x}Mn_{x}As$ results from the interplay between two scales: the 
effective impurity bandwidth of the correlated solution ($W_{eff}$) and the 
local Hund interaction ($J_{H}$) with $J_{H}>W_{eff}$, putting the system 
in the DE class.  As described above, this is out of scope of pure LDA- or 
LDA+U based approaches, which cannot access dynamical effects of strong 
electronic correlations and static (dynamic) disorder. A proper treatment of 
these dynamical processes plays a crucial role  (via DMFT) in generating an 
impurity band with reduced bandwidth ($W<J_{H}$), stabilizing DE 
ferromagnetism in this DMS system.              

\begin{figure}[htb]
\epsfxsize=3.5in
\epsffile{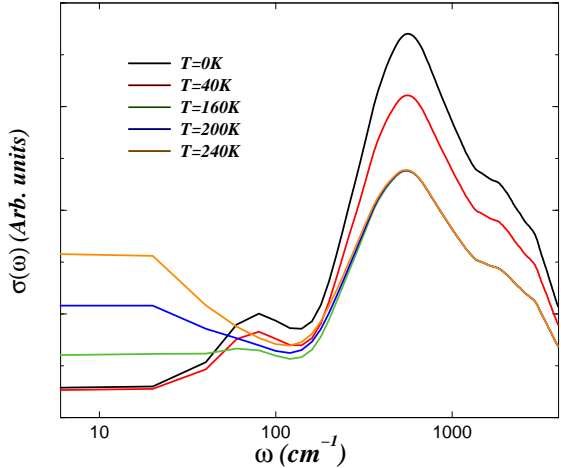}
\caption{Computed optical conductivity $\sigma(\omega,T)$ versus $\omega$ for
various $T$ both below {\it and} above $T_{c}=120~K$.  Good quantitative
agreement with experiment~[9] is clearly visible.} 
\label{fig3}
\end{figure}

Finally, we compute the optical conductivity, $\sigma(\omega)$, directly 
from the fully renormalized DOS using the result known rigorously~\cite{[17]} 
in DMFT.  In Fig.~\ref{fig3}, we show our result, $\sigma(\omega,T)$, as a 
function of $\omega$ for $x=0.043$. Concentrating on the $T$ dependence, we 
see clearly that our results are very similar to the one from Ref.~\cite{[9]}.
In particular, the broad peak around $800~cm^{-1}$, the smaller (anti-site 
related) bump around $100~cm^{-1}$ (somewhat different from experiment), and 
the incoherent low-energy response are all in complete accordance with 
observations. More satisfyingly, the $T$ dependence 
of the spectral weight transfer is also correctly reproduced:  at 
low-$T$ (below $T_{c}$), no crossing point in $\sigma(\omega)$ is seen till 
$3000~cm^{-1}$, while above $T_{c}$, the curves seem to cross around 
$800~cm^{-1}$, in full semiquantitative agreement with experiment.  However, 
we were not able to resolve a clear crossing point in the spectra above 
$T_{c}$. Finally, the distribution and $T$ dependence of the optical spectral 
weight (decrease with increasing $T$ below $T_{c}$, and increasing with $T$ 
above $T_{c}$) is in good agreement with the data as well. 

Given the detailed quantitative agreement of our results with those
gleaned from various spectroscopic, optical, and magnetic measurements, we
believe that we have provided the first totally ab-initio (LDA+correlations)
description of $Ga_{1-x}Mn_{x}As$.  Our study highlights the importance of
including intrinsic {\it and} extrinsic disorder, strong dynamical correlations
and magnetic scattering, in concert with the real LDA bandstructure in one
single picture.  We have clearly shown how a consistent treatment of dynamical 
electronic correlations and disorder generates a narrow, quasicoherent 
impurity band in the semiconductor band gap, opening up the possibility for 
DE to drive the system into a ferromagnetic half-metallic state.  Other 
diluted magnetic semiconductors of great current interest can be studied 
readily within this framework. 

This work is supported by the Sonderforschungsbereich 608 of the Deutsche 
Forschungsgemeinschaft.

\end{document}